 \documentclass[preprint,showpacs,superscriptaddress,nofootinbib]{revtex4}
  \usepackage{graphicx}
  \usepackage{rotating}
  \usepackage{multirow}
  \begin{document}
  \title{$B_{c}$ ${\to}$ $BP$, $BV$ decays with the QCD factorization approach}
  \author{Junfeng Sun}
  \affiliation{Institute of Particle and Nuclear Physics,
              Henan Normal University, Xinxiang 453007, China}
  \author{Na Wang}
  \affiliation{Institute of Particle and Key Laboratory of Quark and Lepton Physics,
              Central China Normal University, Wuhan 430079, China}
  \affiliation{Institute of Particle and Nuclear Physics,
              Henan Normal University, Xinxiang 453007, China}
  \author{Qin Chang}
  \affiliation{Institute of Particle and Nuclear Physics,
              Henan Normal University, Xinxiang 453007, China}
  \author{Yueling Yang}
  \affiliation{Institute of Particle and Nuclear Physics,
              Henan Normal University, Xinxiang 453007, China}
  \begin{abstract}
   In this paper, we studied the nonleptonic $B_{c}$ ${\to}$ $BP$, $BV$
   decay with the QCD factorization approach. It is found that
   the Cabibbo favored processes of $B_{c}$ ${\to}$ $B_{s}{\pi}$,
   $B_{s}{\rho}$, $B_{u}\overline{K}$ are the promising
   decay channels with branching ratio larger than 1\%,
   which should be observed earlier by the LHCb Collaboration.
  \end{abstract}
  \maketitle

  \section{Introduction}
  \label{sec01}
  The $B_{c}$ meson is the ground pseudoscalar meson of the
  $\bar{b}c$ system \cite{pdg}.
  Compared with the heavy unflavored charmonium $c\bar{c}$
  and bottomonium $b\bar{b}$,
  the $B_{c}$ meson is unique in some respects.
  (1)
  Heavy quarkonia could be created in the parton-parton process
  $ij$ ${\to}$ $Q\bar{Q}$ at the order of ${\alpha}_{s}^{2}$
  (where $ij$ $=$ $gg$ or $q\bar{q}$, $Q$ $=$ $b$, $c$),
  while the production probability for the $B_{c}$ meson
  is at least at the order of ${\alpha}_{s}^{4}$ via $ij$
  ${\to}$ $B_{c}^{(\ast)+}$ $+$ $b\bar{c}$ where the gluon-gluon
  fusion mechanism is dominant at Tevatron and LHC \cite{qwg}.
  The $B_{c}$ meson is difficult to produce experimentally,
  but it was observed for the first time via the semileptonic
  decay mode $B_{c}$ ${\to}$ $J/{\psi}{\ell}{\nu}$ in $p\bar{p}$
  collisions by the CDF collaboration in 1998 \cite{1st}, which
  showed the realistic possibility of experimental study of the
  $B_{c}$ meson.
  One of the best measurements on the mass and lifetime
  of the $B_{c}$ meson is reported recently by the LHCb
  collaboration,
  $m_{B_{c}}$ $=$ $6276.28{\pm}1.44{\pm}0.36$ MeV \cite{mass} and
  ${\tau}_{B_{c}}$ $=$ $513.4{\pm}11.0{\pm}5.7$ fs \cite{time}.
  With the running of the LHC, the $B_{c}$ meson has
  a promising prospect.
  It is estimated that one could expect some $10^{10}$ $B_{c}$
  events at the high-luminosity LHC experiments per year \cite{events} .
  The studies on the $B_{c}$ meson have entered a new precision era.
  (2)
  For charmonium and bottomonium, the strong and electromagnetic
  interactions are mainly responsible for annihilation of the
  $Q\bar{Q}$ quark pair into final states.
  The $B_{c}$ meson, carrying nonzero flavor number $B$ $=$ $C$ $=$ ${\pm}1$
  and lying below the $BD$ meson pair threshold, can decay only via the weak
  interaction, which offers an ideal sample to investigate the
  weak decay mechanism of heavy flavors that is inaccessible to
  both charmonium and bottomonium.
  The $B_{c}$ weak decay provides great opportunities to
  investigate the perturbative and nonperturbative QCD, final
  state interactions, etc.

  With respect to the heavy-light $B_{u,d,s}$ mesons, the doubly heavy
  $B_{c}$ meson has rich decay channels because of its relatively large
  mass and that both $b$ and $c$ quarks can decay individually.
  The decay processes of the $B_{c}$ meson can be divided into
  three classes \cite{qwg,zpc51,prd49,usp38}:
  (1) the $c$ quark decays with the $b$ quark as a spectator;
  (2) the $b$ quark decays with the $c$ quark as a spectator;
  (3) the $b$ and $c$ quarks annihilate into a virtual $W$ boson,
  with the ratios of ${\sim}$ 70\%, 20\% and 10\%, respectively
  \cite{qwg}.
  Up to now, the experimental evidences of pure annihilation
  decay mode [class (3)] are still nothing.
  The $b$ ${\to}$ $c$ transition, belonging to the class (2),
  offers a well-constructed experimental
  structure of charmonium at the Tevatron and LHC.
  Although the detection of the $c$ quark decay is very
  challenging to experimentalists, the clear signal of
  the $B_{c}$ ${\to}$ $B_{s}{\pi}$ decay is presented by
  the LHCb group using the 
  $B_{s}$ ${\to}$ $D_{s}{\pi}$
  and $B_{s}$ ${\to}$ $J/{\psi}{\phi}$ channels with
  statistical significance of $7.7{\sigma}$ and
  $6.1{\sigma}$, respectively \cite{bspi}.

  Anticipating the forthcoming accurate measurements
  on the $B_{c}$ meson at hadron colliders
  and the lion's share of the $B_{c}$ decay width
  from the $c$ quark decay \cite{prd53},
  many theoretical papers have devoted to the study
  of the $B_{c}$ ${\to}$ $BP$, $BV$ decays (where $P$
  and $V$ denote the $SU(3)$ ground pseudoscalar and
  vector mesons, respectively),
  such as Refs. \cite{prd.39.1342,zpc.51.549,pan.62.1793,jpg.35.085002}
  with the BSW model \cite{bsw} or IGSW model \cite{igsw},
  Refs. \cite{prd.49.3399,prd.56.4133,prd.62.014019}
  based on the Bethe-Salpeter (BS) equation,
  Refs. \cite{prd.61.034012,epjc.32.29,prd.74.074008,prd.80.114003,prd.86.094028}
  with potential models,
  Ref. \cite{prd.73.054024} with constituent quark model,
  Ref. \cite{pan.67.1559} with QCD sum rules,
  Ref. \cite{prd.65.114007} with the quark diagram scheme,
  Ref. \cite{prd.89.114019} with the perturbative QCD approach (pQCD)
  \cite{pqcd}, and so on.
  The previous predictions on the branching ratios for
  the $B_{c}$ ${\to}$ $BP$, $BV$ decays are collected in
  Table \ref{tab04}.
  The discrepancies of previous investigations
  arise mainly from the different model assumptions.
  Recently, several phenomenological methods have been
  fully developed to cope with the hadronic matrix
  elements and successfully applied to the nonleptonic $B$
  decay, such as the pQCD approach \cite{pqcd}
  based on the $k_{T}$ factorization scheme,
  the soft-collinear effective theory \cite{scet} and
  the QCD-improved factorization (QCDF) approach \cite{qcdf}
  based on the collinear approximation and power countering
  rules in the heavy quark limits.
  In this paper, we will study the
  $B_{c}$ ${\to}$ $BP$, $BV$ decay with the QCDF approach
  to provide a ready reference to the existing and
  and upcoming experiments.

  This paper is organized as follows.
  In section \ref{sec02}, we will present the theoretical framework
  and the amplitudes for the $B_{c}$ ${\to}$ $BP$, $BV$
  decays within the QCDF framework.
  The section \ref{sec03} is devoted to numerical results and discussion.
  Finally, the section \ref{sec04} is our summation.

  \section{theoretical framework}
  \label{sec02}
  \subsection{The effective Hamiltonian}
  \label{sec0201}
  The low energy effective Hamiltonian responsible for the nonleptonic
  bottom-conserving $B_{c}$ ${\to}$ $BP$, $BV$ decays constructed by
  means of the operator product expansion and the renormalization
  group (RG) method is usually written in terms of the
  four-quark interactions \cite{9512380}.
   \begin{eqnarray}
   H_{\rm eff}
  &=&
   \frac{G_{F}}{\sqrt{2}} \Big\{
   V_{ub} V_{cb}^{\ast}\, [
   C_{1}^{a}({\mu})\,Q_{1}^{a}({\mu})
  +C_{2}^{a}({\mu})\,Q_{2}^{a}({\mu}) ]
   \nonumber \\ & &
  +\sum\limits_{q_{1},q_{2}=d,s}
   V_{uq_{1}} V_{cq_{2}}^{\ast}\, [
   C_{1}({\mu})\,Q_{1}({\mu})
  +C_{2}({\mu})\,Q_{2}({\mu}) ]
   \nonumber \\ & &
  +\sum\limits_{q_{3}=d,s}
   V_{uq_{3}} V_{cq_{3}}^{\ast}\,
   \sum\limits_{k=3}^{10}
   C_{k}({\mu})\,Q_{k}({\mu})
   \Big\} + {\rm h.c.}
   \label{e01},
   \end{eqnarray}
  where the Fermi coupling constant $G_{F}$ ${\simeq}$ $1.166$
  ${\times}$ $10^{-5}$ ${\rm GeV}^{-2}$ \cite{pdg};
  $Q_{1,2}$, $Q_{1,2}^{a}$ and $Q_{3,{\cdots}10}$ are the relevant
  local tree, annihilation and penguin four-quark operators,
  respectively, which govern the decays in question.
  The Cabibbo-Kobayashi-Maskawa (CKM) factor
  $V_{uq_{i}} V_{cq_{j}}^{\ast}$ and Wilson coefficients $C_{i}$
  describe the coupling strength for a given operator.

  Using the unitarity of the CKM matrix, there is a large
  cancellation of the CKM factors
  \begin{equation}
     V_{ud}V_{cd}^{\ast}
    +V_{us}V_{cs}^{\ast}
   =-V_{ub}V_{cb}^{\ast}\
   {\sim}\ {\cal O}({\lambda}^{5})
  \label{ckm},
  \end{equation}
  where the Wolfenstein parameter ${\lambda}$
  $=$ ${\sin}{\theta}_{c}$ $=$ $0.225\, 37(61)$
  \cite{pdg} and ${\theta}_{c}$ is the Cabibbo angle.
  Hence, compared with the tree contributions,
  the contributions of annihilation and penguin operators
  are strongly suppressed by the CKM factor.
  If the $CP$-violating asymmetries that are expected
  to be very small due to the small weak phase difference
  for $c$ quark decay are prescinded from the present
  consideration, then the penguin and annihilation
  contributions could be safely neglected.
  The local tree operators $Q_{1,2}$ in Eq.(\ref{e01})
  are expressed as follows,
  \begin{eqnarray}
  Q_{1} &=&
  [ \bar{q}_{2,{\alpha}}{\gamma}_{\mu}(1-{\gamma}_{5})c_{\alpha} ]
  [ \bar{u}_{\beta}{\gamma}^{\mu}(1-{\gamma}_{5})q_{1,{\beta}} ]
  \label{q1}, \\
  Q_{2} &=&
  [ \bar{q}_{2,{\alpha}}{\gamma}_{\mu}(1-{\gamma}_{5})c_{\beta} ]
  [ \bar{u}_{\beta}{\gamma}^{\mu}(1-{\gamma}_{5})q_{1,{\alpha}} ]
  \label{q2},
  \end{eqnarray}
  where ${\alpha}$ and ${\beta}$ are the $SU(3)$ color indices.

  The Wilson coefficients $C_{i}({\mu})$ summarize the physics
  contributions from scales higher than ${\mu}$.
  They are calculable with the RG improved
  perturbation theory and have properly been evaluated to
  the next-to-leading order (NLO) \cite{9512380}.
  They can be evolved from a higher scale
  ${\mu}$ ${\sim}$ ${\cal O}(m_{W})$
  down to a characteristic scale
  ${\mu}$ ${\sim}$ ${\cal O}(m_{c})$
  with the functions including the flavor thresholds \cite{9512380},
  \begin{equation}
  \vec{C}({\mu}) = U_{4}({\mu},m_{b})M(m_{b})U_{5}(m_{b},m_{W})\vec{C}(m_{W})
  \label{ci},
  \end{equation}
  where $U_{f}({\mu}_{f},{\mu}_{i})$ is the RG evolution matrix
  converting coefficients from the scale ${\mu}_{i}$ to ${\mu}_{f}$,
  and $M({\mu})$ is the quark threshold matching matrix.
  The expressions of
  $U_{f}({\mu}_{f},{\mu}_{i})$ and $M({\mu})$
  can be found in Ref. \cite{9512380}.
  The numerical values of LO and NLO $C_{1,2}$ with the
  naive dimensional regularization scheme are listed
  in Table \ref{tab02}.
  The values of NLO Wilson coefficients in Table \ref{tab02}
  are consistent with those given by Ref. \cite{9512380}
  where a trick with ``effective'' number of active flavors
  $f$ $=$ 4.15 rather than formula Eq.(\ref{ci}) is used.

  To obtain the decay amplitudes, the remaining work is
  how to accurately evaluate the hadronic matrix elements
  ${\langle}BM{\vert}Q_{i}({\mu}){\vert}B_{c}{\rangle}$
  which summarize the physics contributions from scales
  lower than ${\mu}$.
  Since the hadronic matrix elements involve long distance
  contributions, one is forced to use either
  nonperturbative methods such as lattice calculations
  and QCD sum rules or phenomenological models relying on
  some assumptions.
  Consequently, it is very unfortunate that hadronic matrix
  elements cannot be reliably calculated at present, and that
  the most intricate part and the dominant theoretical
  uncertainties in the decay amplitudes reside in the
  hadronic matrix elements.

  \subsection{Hadronic matrix elements}
  \label{sec0202}
  Phenomenologically,
  based on the power counting rules in the heavy quark limit,
  M. Beneke {\em et al}. proposed that the hadronic matrix
  elements could be written as the convolution integrals
  of hard scattering kernels and the light cone distribution
  amplitudes with the QCDF master formula \cite{qcdf}.
  The QCDF approach is widely applied to nonleptonic $B$ decays
  and it works well \cite{du,beneke,prd.73.114027,cheng},
  which encourage us to apply the QCDF approach to the study of
  $B_{c}$ ${\to}$ $BP$, $BV$ decay.
  Since the spectator is the heavy $b$ quark who is almost always
  on shell, the virtuality of the gluon linked with the spectator
  should be  ${\sim}$ ${\cal O}({\Lambda}^{2}_{\rm QCD})$.
  The dominant behavior of the $B_{c}$ ${\to}$ $B$ transition
  form factors and the contributions of hard spectator scattering
  interactions are governed by soft processes.
  According to the basic idea of the QCDF approach \cite{beneke},
  the hard and soft contributions to the form factors entangle
  with each other and cannot be identified reasonably, so the
  physical form factors are used as the inputs.
  The hard spectator scattering contributions
  are power suppressed in the heavy quark limit.
  Finally, the hadronic matrix elements can be written as,
  \begin{eqnarray}
  {\langle}BM{\vert}Q_{1,2}{\vert}B_{c}{\rangle}
  &=&
   \sum\limits_{i} F_{i}^{ B_{c}{\to}B }
  {\int}\,dx\, H_{i}(x)\,{\Phi}_{M}(x)
   \nonumber \\
  &{\propto}&
  \sum\limits_{i} F_{i}^{ B_{c}{\to}B }f_{M}
  \{1+{\alpha}_{s}r_{1}+{\cdots}\}
  \label{hadronic},
  \end{eqnarray}
  where $F_{i}^{ B_{c}{\to}B }$ is the transition form factor and
  ${\Phi}_{M}(x)$ is the light-cone distribution amplitudes of
  the emitted meson $M$ with the decay constant $f_{M}$.
  The hard scattering kernels $H_{i}(x)$ are computable order by
  order with the perturbation theory in principle.
  At the leading order ${\alpha}_{s}^{0}$, $H_{i}(x)$ $=$ $1$,
  i.e., the convolution integral of Eq.(\ref{hadronic})
  results in the meson decay constant.
  The hadronic matrix elements are parameterized by the
  product of form factors and decay constants, which
  are real and renormalization scale independent.
  One goes back to the simple ``naive factorization'' (NF) scenario.
  At the order ${\alpha}_{s}$ and higher orders,
  the information of strong phases and the renormalization
  scale dependence of hadronic matrix elements could be
  partly recuperated.
  Combined the nonfactorizable contributions with the Wilson coefficients,
  the scale independent effective coefficients at the order ${\alpha}_{s}$
  can be obtained \cite{qcdf} as follows:
  \begin{eqnarray}
   a_{1}
   &=& C_{1}^{\rm NLO}+\frac{1}{N_{c}}\,C_{2}^{\rm NLO}
    + \frac{{\alpha}_{s}}{4{\pi}}\, \frac{C_{F}}{N_{c}}\,
      C_{2}^{\rm LO}\, V
   \label{a1}, \\
   a_{2}
   &=& C_{2}^{\rm NLO}+\frac{1}{N_{c}}\,C_{1}^{\rm NLO}
    + \frac{{\alpha}_{s}}{4{\pi}}\, \frac{C_{F}}{N_{c}}\,
      C_{1}^{\rm LO}\, V
   \label{a2},
  \end{eqnarray}
  where the expression of vertex corrections are \cite{qcdf}:
  \begin{equation}
  V = 6\,{\log}\Big( \frac{m_{c}^{2}}{{\mu}^{2}} \Big)
    -  18 - \Big( \frac{1}{2}+i3{\pi} \Big)\,a_{0}^{M}
    +  \Big( \frac{11}{2}-i3{\pi} \Big)\,a_{1}^{M}
    -   \frac{21}{20}\,a_{2}^{M} +{\cdots}
  \label{vc},
  \end{equation}
  with the twist-2 quark-antiquark distribution amplitudes of
  pseudoscalar $P$ and longitudinally polarized vector $V$ meson
  in terms of Gegenbauer polynomials \cite{ball}:
   \begin{equation}
  {\phi}_{M}(x)=6\,x\bar{x}
   \sum\limits_{n=0}^{\infty}
   a_{n}^{M}\, C_{n}^{3/2}(x-\bar{x})
   \label{twist},
   \end{equation}
  where $\bar{x}$ $=$ $1$ $-$ $x$;
  $a_{n}^{M}$ is the Gegenbauer moment
  and $a_{0}^{M}$ ${\equiv}$ $1$.

  From the numbers in Table \ref{tab02}, it is found that
  (1) for the coefficient $a_{1}$, the nonfactorizable
  contributions accompanied by the Wilson coefficient
  $C_{2}$ can provide ${\ge}$ 10\% enhancement compared
  with the NF's result, and a relatively small strong
  phase ${\le}$ $5^{\circ}$;
  (2) for the coefficient $a_{2}$, the nonfactorizable
  contributions assisted with the large Wilson coefficient
  $C_{1}$ are significant.
  In addition, a relatively large strong phase ${\sim}$
  $-155^{\circ}$ is obtained;
  (3) the QCDF's values of $a_{1,2}$ agree basically with
  the real coefficients $a_{1}$ ${\simeq}$ $1.20$ and
  $a_{2}$ ${\simeq}$ $-0.317$ which are
  used by previous studies on the $B_{c}$ ${\to}$ $BP$, $BV$
  decays in Refs. \cite{prd.39.1342,zpc.51.549,pan.62.1793,
  jpg.35.085002,prd.49.3399,
  prd.61.034012,epjc.32.29,prd.74.074008,prd.80.114003,
  prd.86.094028,prd.73.054024,pan.67.1559},
  but with more information on the strong phases.

  \subsection{Decay amplitudes}
  \label{sec0203}
  Within the QCDF framework, the amplitudes
  for $B_{c}$ ${\to}$ $BM$ decays are expressed as:
   \begin{equation}
  {\cal A}(B_{c}{\to}BM)\ =\
  {\langle}BM{\vert}{\cal H}_{\rm eff}{\vert}B_{c}{\rangle}\ =\
   \frac{G_{F}}{\sqrt{2}}\,
   V_{uq_{1}} V_{cq_{2}}^{\ast}\, a_{i}\,
  {\langle}M{\vert}J^{\mu}{\vert}0{\rangle}
  {\langle}B{\vert}J_{\mu}{\vert}B_{c}{\rangle}
   \label{lorentz}.
   \end{equation}

  The matrix elements of current operators are defined as:
   \begin{eqnarray}
  {\langle}P(p){\vert}\bar{q}_{1}{\gamma}^{\mu}(1-{\gamma}_{5})q_{2}
  {\vert}0{\rangle}
  &=&
  -i\,f_{P}\,p^{\mu}
   \label{pseudoscalar}, \\
  {\langle}V({\epsilon},p){\vert}\bar{q}_{1}{\gamma}^{\mu}(1-{\gamma}_{5})q_{2}
  {\vert}0{\rangle}
  &=&
  f_{V}\,m_{V}{\epsilon}^{\mu}
   \label{vector}.
   \end{eqnarray}
  where $f_{P}$ and $f_{V}$ are the decay constants
  of pseudoscalar $P$ and vector $V$ mesons, respectively;
  $m_{V}$ and ${\epsilon}$ denote the mass and
  polarization of vector meson, respectively.

  For the mixing of physical pseudoscalar ${\eta}$ and ${\eta}^{\prime}$
  meson, we adopt the quark-flavor basis description proposed
  in Ref. \cite{prd.58.114006}, and neglect the contributions
  from possible gluonium and $c\bar{c}$ compositions, i.e.,
   \begin{equation}
   \left(\begin{array}{c}
  {\eta} \\ {\eta}^{\prime}
   \end{array}\right) =
   \left(\begin{array}{cc}
  {\cos}{\phi} & -{\sin}{\phi} \\
  {\sin}{\phi} &  {\cos}{\phi}
   \end{array}\right)
   \left(\begin{array}{c}
  {\eta}_{q} \\ {\eta}_{s}
   \end{array}\right)
   \label{mixing01},
   \end{equation}
  where ${\eta}_{q}$ $=$ $(u\bar{u}+d\bar{d})/{\sqrt{2}}$
  and ${\eta}_{s}$ $=$ $s\bar{s}$;
  the mixing angle ${\phi}$ $=$ $(39.3{\pm}1.0)^{\circ}$
  \cite{prd.58.114006}.
  We assume that the vector mesons are ideally mixed,
  i.e., ${\omega}$ $=$ $(u\bar{u}+d\bar{d})/\sqrt{2}$
  and ${\phi}$ $=$ $s\bar{s}$.

  The transition form factors are defined as \cite{bsw}:
   \begin{eqnarray}
  & &
  {\langle}B(k){\vert}\bar{q}{\gamma}^{\mu}(1-{\gamma}_{5})c
  {\vert}B_{c}(p){\rangle}
  \nonumber \\ &=&
  \Big[ p+k - \frac{m_{B_{c}}^{2}-m_{B}^{2}}{q^{2}}q \Big]^{\mu}
  F_{1}^{B_{c}{\to}B}(q^{2})
  \nonumber \\ &+&
  \frac{m_{B_{c}}^{2}-m_{B}^{2}}{q^{2}}q^{\mu}
  F_{0}^{B_{c}{\to}B}(q^{2})
   \label{ff01},
   \end{eqnarray}
  where $q$ $=$ $p$ $-$ $k$,
  and the condition of
  $F_{0}^{B_{c}{\to}B}(0)$ $=$ $F_{1}^{B_{c}{\to}B}(0)$
  is required compulsorily to cancel the singularity
  at the pole $q^{2}$ $=$ $0$.

  For the $B_{c}$ ${\to}$ $B$ transition form factors,
  since the velocity of the recoiled $B$ meson is very
  low in the rest frame of the $B_{c}$ meson, the wave
  functions of the $B$ and $B_{c}$ meson overlap strongly.
  It is believed that the form factors
  $F_{0,1}^{B_{c}{\to}B}$ should be close to the result
  using the nonrelativistic harmonic oscillator wave
  functions with the BSW model \cite{prd.39.1342},
  \begin{equation}
  F_{0,1}^{B_{c}{\to}B}\ {\simeq}\
  \left( \frac{ 2\,m_{B_{c}}m_{B} }{ m_{B_{c}}^{2}+m_{B}^{2} }
  \right)^{1/2}\ {\simeq}\
  0.99
  \label{ff02}.
  \end{equation}
  The flavor symmetry breaking effects on the
  form factors are neglectable in Eq.(\ref{ff02}).
  For simplification, we take $F_{0,1}^{B_{c}{\to}B_{u,d,s}}$ $=$ 1.0
  in our numerical calculation to give a rough estimation.

  \section{numerical results and discussions}
  \label{sec03}
  The branching ratios of nonleptonic two-body $B_{c}$ decays
  in the rest frame of the $B_{c}$ meson can be written as:
  \begin{equation}
  {\cal B}r(B_{c}{\to}BM)=
   \frac{{\tau}_{B_{c}}}{8{\pi}}
   \frac{p}{m_{B_{c}}^{2}}
  {\vert}{\cal A}(B_{c}{\to}BM){\vert}^{2}
   \label{br01},
   \end{equation}
  where the lifetime of the $B_{c}$ meson ${\tau}_{B_{c}}$ $=$
  $513.4{\pm}11.0{\pm}5.7$ fs \cite{time} and $p$ is the common
  momentum of final particles.
  The decay amplitudes ${\cal A}(B_{c}{\to}BM)$
  are listed in Appendix \ref{app04}.

  The input parameters in our calculation, including the CKM
  Wolfenstein parameters, decay constants, Gegenbauer
  moments of distribution amplitudes in Eq.(\ref{twist}),
  are collected in Table \ref{input}.
  If not specified explicitly, we will take their central values
  as the default inputs.
  Our numerical results on the $CP$-averaged branching ratios
  are presented in Table \ref{tab04}, where theoretical
  uncertainties of the ``QCDF'' column come from the CKM parameters,
  the renormalization scale ${\mu}$ $=$ $(1{\pm}0.2)m_{c}$,
  decay constants and Gegenbauer moments, respectively.
  For comparison, previous results calculated with the
  fixed coefficients $a_{1}$ ${\simeq}$ $1.22$ and $a_{2}$ ${\simeq}$
  $-0.4$ are also listed.
  There are some comments on the branching ratios.

  (1)
  From the numbers in Table \ref{tab04}, it is seen that
  different branching ratios for the $B_{c}$ ${\to}$ $BP$, $BV$
  decays were obtained with different approaches in previous works,
  although the same coefficients $a_{1,2}$ are used.
  Much of the discrepancy comes from the different values of
  the transition form factors. If the same value of the form
  factor is used, then the disparities on branching ratios
  for the $a_{1}$-dominated $B_{c}$ decays will be highly
  alleviated.
  For example, all previous predictions on
  ${\cal B}r(B_{c}{\to}B_{s}{\pi})$
  will be about $10\%$ with the same form factor
  $F_{0}^{B_{c}{\to}B_{s}}$ ${\simeq}$ $1.0$,
  which is generally in line with the QCDF estimation within
  uncertainties and also agrees with the recent LHCb
  measurement \cite{bspi}.

  (2)
  There is a hierarchical structure between the
  QCDF's results on branching ratios for the $B_{c}$
  ${\to}$ $BP$ and $BV$ decays with the same final
  $B_{q}$ meson, for example,
   \begin{eqnarray}
  {\cal B}r(B_{c}{\to}B_{q}{\pi}) >
  {\cal B}r(B_{c}{\to}B_{q}{\rho})
   \label{pirho}, \\
  {\cal B}r(B_{c}{\to}B_{q}K)\ {\gtrsim}\ 5\,
  {\cal B}r(B_{c}{\to}B_{q}K^{\ast})
   \label{kk},
   \end{eqnarray}
  which differs from the previous results.
  There are two decisive factors.
  One is the kinematic factor.
  The phase space for the $B_{c}$ ${\to}$ $BV$ decays
  is more compressed than that for the $B_{c}$ ${\to}$
  $BP$ decays, because the mass of the light pseudoscalar
  meson (except for the exotic ${\eta}^{\prime}$ meson)
  is generally less than the mass of the corresponding
  vector meson with the same valence quark components.
  The other is the dynamical factor.
  The orbital angular momentum for the $BP$ final states
  is ${\ell}_{BP}$ $=$ $0$, while the orbital angular
  momentum ${\ell}_{BV}$ $=$ $1$ for the $BV$ final states.

  (3)
  According to the CKM factors and the coefficients $a_{1,2}$,
  there is another hierarchy of amplitudes among the QCDF's
  branching ratios for the $B_{c}$ decays, which could be
  subdivided into different cases as below.
  The CKM-favored $a_{1}$-dominated $B_{c}$ ${\to}$
  $B_{s}{\pi}$ decay are expected to have the largest branching
  ratio, ${\sim}$ 10\%, within the QCDF framework.
  In addition, the branching ratios for the Cabibbo favored
  $B_{c}^{+}$ ${\to}$ $B_{s}^{0}{\rho}^{+}$,
  $B_{u}^{+}\overline{K}^{0}$ decays are
  also larger than 1\%, which might be promisingly detected
  at experiments.
  \begin{center}
  \begin{ruledtabular}
  \begin{tabular}{ccccc}
  Case & Coefficient & CKM factor & Branching ratio & Decay modes\\ \hline
  1 a  & $a_{1}$ & ${\vert}V_{ud}V_{cs}^{\ast}{\vert}$ ${\sim}$ $1$
                 & ${\gtrsim}$ $10^{-2}$
                 & $B_{s}{\pi}$, $B_{s}{\rho}$ \\
  1 b  & $a_{1}$ & ${\vert}V_{ud}V_{cd}^{\ast}{\vert}$,
                   ${\vert}V_{us}V_{cs}^{\ast}{\vert}$ ${\sim}$ ${\lambda}$
                 & ${\gtrsim}$ $10^{-3}$
                 & $B_{s}K$, $B_{d}{\pi}$, $B_{d}{\rho}$ \\
  1 c  & $a_{1}$ & ${\vert}V_{us}V_{cd}^{\ast}{\vert}$ ${\sim}$ ${\lambda}^{2}$
                 & ${\gtrsim}$ $10^{-5}$
                 & $B_{d}K$, $B_{d}K^{\ast}$  \\ \hline
  2 a  & $a_{2}$ & ${\vert}V_{ud}V_{cs}^{\ast}{\vert}$ ${\sim}$ $1$
                 & ${\gtrsim}$ $10^{-3}$
                 & $B_{u}^{+}\overline{K}^{0}$, $B_{u}^{+}\overline{K}^{{\ast}0}$ \\
  2 b  & $a_{2}$ & ${\vert}V_{ud}V_{cd}^{\ast}{\vert}$,
                   ${\vert}V_{us}V_{cs}^{\ast}{\vert}$ ${\sim}$ ${\lambda}$
                 & ${\gtrsim}$ $10^{-4}$
                 & $B_{u}{\pi}$, $B_{u}{\rho}$, $B_{u}{\omega}$ \\
  2 c  & $a_{2}$ & ${\vert}V_{us}V_{cd}^{\ast}{\vert}$ ${\sim}$ ${\lambda}^{2}$
                 & ${\gtrsim}$ $10^{-6}$
                 & $B_{u}^{+}K^{0}$, $B_{u}^{+}K^{{\ast}0}$
  \end{tabular}
  \end{ruledtabular}
  \end{center}

  (4)
  There are many uncertainties on the QCDF's results.
  The first uncertainty from the CKM factors is small due to the
  high precision on Wolfenstein parameter ${\lambda}$ with only
  0.3\% relative errors \cite{pdg}.
  Large uncertainty comes from the renormalization scale,
  especially for the $a_{2}$ dominated $B_{c}$ ${\to}$
  $B_{u}P$, $B_{u}V$ decays.
  In principle, the second uncertainty could be reduced by the
  inclusion of higher order ${\alpha}_{s}$ corrections to
  hadronic matrix elements.
  It has been showed \cite{scale} that tree amplitudes
  incorporating with the NNLO corrections are relatively
  less sensitive to the choice of scale than the NLO amplitudes.
  As aforementioned, large uncertainty mainly comes from hadron
  parameters, such as the transition form factors, which
  is expected to be cancelled from the rate of branching ratios.
  For example,
   \begin{eqnarray}
   & &
   \frac{ {\cal B}r(B_{c}{\to}B_{s}K) }{ {\cal B}r(B_{c}{\to}B_{s}{\pi}) }
   \ {\approx}\ {\vert}V_{us}{\vert}^{2}\frac{ f_{K}^{2} }{ f_{\pi}^{2} }
   \ {\approx}\
   \frac{ {\cal B}r(B_{c}{\to}B_{d}K) }{ {\cal B}r(B_{c}{\to}B_{d}{\pi}) }
   \label{r02}, \\
  & &
   \frac{ {\cal B}r(B_{c}{\to}B_{s}K^{\ast}) }{ {\cal B}r(B_{c}{\to}B_{s}{\rho}) }
   \ {\approx}\ {\vert}V_{us}{\vert}^{2}\frac{ f_{K^{\ast}}^{2} }{ f_{\rho}^{2} }
   \ {\approx}\
   \frac{ {\cal B}r(B_{c}{\to}B_{d}K^{\ast}) }{ {\cal B}r(B_{c}{\to}B_{d}{\rho}) }
   \label{r04}, \\
   & &
   \frac{ {\cal B}r(B_{c}{\to}B_{u}{\pi}) }{ {\cal B}r(B_{c}{\to}B_{d}{\pi}) }
   \ {\approx}\ \frac{1}{2} \frac{ {\vert}a_{2}{\vert}^{2} }{ {\vert}a_{1}{\vert}^{2} }
   \ {\approx}\
   \frac{ {\cal B}r(B_{c}{\to}B_{u}{\rho}) }{ {\cal B}r(B_{c}{\to}B_{d}{\rho}) }
   \label{r06}.
   \end{eqnarray}
  Especially, the relation of Eq.(\ref{r06}) might be used to give
  some information on the coefficients $a_{1,2}$ and to provide an interesting
  feasibility research on the validity of the QCDF approach
  for the charm quark decay.
  Finally, we would like to point out that many uncertainties
  from other factors, such as the final state interactions,
  which deserve the dedicated study, are not considered here.
  So one should not be too serious about the numbers in
  Table \ref{tab04}.
  Despite this, our results will still provide some useful
  information to experimental physicists, i.e.,
  the Cabibbo favored $B_{c}$ ${\to}$ $B_{s}{\pi}$, $B_{s}{\rho}$,
  $B_{u}\overline{K}$ decays have large branching ratios ${\gtrsim}$
  1\%, which could be detected earlier.

  \section{Summary}
  \label{sec04}
  In prospects of the potential $B_{c}$ meson at the LHCb experiments,
  accurate and thorough studies of the $B_{c}$ decays will be
  accessible very soon.
  The carefully theoretical study on the $B_{c}$ decays is
  urgently desiderated.
  In this paper, we concentrated on the nonfactorizable contributions
  to hadronic matrix elements within the QCDF framework, while the
  transition form factors are taken as nonperturbative inputs,
  which is different from previous studies.
  It is found that the branching ratios for the Cabibbo favored
  $B_{c}$ ${\to}$ $B_{s}{\pi}$, $B_{s}{\rho}$, $B_{u}\overline{K}$
  decays are very large and could be measured earlier by the running
  LHCb experiment in the forthcoming years.

  \section*{Acknowledgments}
  The work is supported by the National Natural Science Foundation
  of China (Grant Nos. 11475055, 11275057 and U1232101).
  N. Wang thanks for the support from CCNU-QLPL Innovation Fund (QLPL201411).
  Q. Chang is also supported by the Foundation for the Author of National
  Excellent Doctoral Dissertation of P. R. China (Grant No. 201317)
  and the Program for Science and Technology Innovation Talents in
  Universities of Henan Province (Grant No. 14HASTIT036).
  \begin{appendix}
  \section{decay amplitudes}
  \label{app04}
   \begin{eqnarray}
  {\cal A}(B_{c}^{+}{\to}B_{s}^{0}{\pi}^{+})
   &=&
   -i \frac{G_{F}}{\sqrt{2}}\, F^{B_{c}{\to}B_{s}}_{0}\, f_{\pi}\,
   (m_{B_{c}}^{2}-m_{B_{s}}^{2})\, V_{ud}V_{cs}^{\ast}\, a_{1}
   \label{amp-bspi}, \\
  {\cal A}(B_{c}^{+}{\to}B_{s}^{0}K^{+})
   &=&
   -i \frac{G_{F}}{\sqrt{2}}\, F^{B_{c}{\to}B_{s}}_{0}\, f_{K}\,
   (m_{B_{c}}^{2}-m_{B_{s}}^{2})\, V_{us}V_{cs}^{\ast}\, a_{1}
   \label{amp-bskp}, \\
  {\cal A}(B_{c}^{+}{\to}B_{s}^{0}{\rho}^{+})
   &=&
   \sqrt{2}\, G_{F}\, F^{B_{c}{\to}B_{s}}_{1}\, f_{\rho}\, m_{\rho}\,
   ({\epsilon}_{\rho}{\cdot}p_{B_{c}})\, V_{ud}V_{cs}^{\ast}\, a_{1}
   \label{amp-bsrho}, \\
  {\cal A}(B_{c}^{+}{\to}B_{s}^{0}K^{{\ast}+})
   &=&
   \sqrt{2}\, G_{F}\, F^{B_{c}{\to}B_{s}}_{1}\, f_{K^{\ast}}\, m_{K^{\ast}}\,
   ({\epsilon}_{K^{\ast}}{\cdot}p_{B_{c}})\, V_{us}V_{cs}^{\ast}\, a_{1}
   \label{amp-bskv}, \\
  {\cal A}(B_{c}^{+}{\to}B_{d}^{0}{\pi}^{+})
   &=&
   -i \frac{G_{F}}{\sqrt{2}}\, F^{B_{c}{\to}B_{d}}_{0}\, f_{\pi}\,
   (m_{B_{c}}^{2}-m_{B_{d}}^{2})\, V_{ud}V_{cd}^{\ast}\, a_{1}
   \label{amp-bdpi}, \\
  {\cal A}(B_{c}^{+}{\to}B_{d}^{0}K^{+})
  &=&
  -i \frac{G_{F}}{\sqrt{2}}\, F^{B_{c}{\to}B_{d}}_{0}\, f_{K}\,
   (m_{B_{c}}^{2}-m_{B_{d}}^{2})\, V_{us}V_{cd}^{\ast}\, a_{1}
   \label{amp-bdkp}, \\
  {\cal A}(B_{c}^{+}{\to}B_{d}^{0}{\rho}^{+})
   &=&
   \sqrt{2}\, G_{F}\, F^{B_{c}{\to}B_{d}}_{1}\, f_{\rho}\, m_{\rho}
   ({\epsilon}_{\rho}{\cdot}p_{B_{c}})\, V_{ud}V_{cd}^{\ast}\, a_{1}
   \label{amp-bdrho}, \\
  {\cal A}(B_{c}^{+}{\to}B_{d}^{0}K^{{\ast}+})
   &=&
   \sqrt{2}\, G_{F}\, F^{B_{c}{\to}B_{d}}_{1}\, f_{K^{\ast}}\, m_{K^{\ast}}\,
   ({\epsilon}_{K^{\ast}}{\cdot}p_{B_{c}})\, V_{us}V_{cd}^{\ast}\, a_{1}
   \label{amp-bdkv}, \\
  {\cal A}(B_{c}^{+}{\to}B_{u}^{+}\overline{K}^{0})
   &=&
   -i \frac{G_{F}}{\sqrt{2}}\, F^{B_{c}{\to}B_{u}}_{0}\, f_{K}\,
   (m_{B_{c}}^{2}-m_{B_{u}}^{2})\, V_{ud}V_{cs}^{\ast}\, a_{2}
   \label{amp-bukb}, \\
  {\cal A}(B_{c}^{+}{\to}B_{u}^{+}K^{0})
   &=&
   -i \frac{G_{F}}{\sqrt{2}}\, F^{B_{c}{\to}B_{u}}_{0}\, f_{K}\,
   (m_{B_{c}}^{2}-m_{B_{u}}^{2})\, V_{us}V_{cd}^{\ast}\, a_{2}
   \label{amp-bukz}, \\
  {\cal A}(B_{c}^{+}{\to}B_{u}^{+}\overline{K}^{{\ast}0})
   &=&
   \sqrt{2}\, G_{F}\, F^{B_{c}{\to}B_{u}}_{1}\, f_{K^{\ast}}\, m_{K^{\ast}}\,
   ({\epsilon}_{K^{\ast}}{\cdot}p_{B_{c}})\, V_{ud}V_{cs}^{\ast}\, a_{2}
   \label{amp-bukvb}, \\
  {\cal A}(B_{c}^{+}{\to}B_{u}^{+}K^{{\ast}0})
   &=&
   \sqrt{2}\, G_{F}\, F^{B_{c}{\to}B_{u}}_{1}\, f_{K^{\ast}}\, m_{K^{\ast}}\,
   ({\epsilon}_{K^{\ast}}{\cdot}p_{B_{c}})\, V_{us}V_{cd}^{\ast}\, a_{2}
   \label{amp-bukvz}, \\
   {\cal A}(B_{c}^{+}{\to}B_{u}^{+}{\pi}^{0})
   &=&
   +i \frac{G_{F}}{2}\, F^{B_{c}{\to}B_{u}}_{0}\, f_{\pi}\,
   (m_{B_{c}}^{2}-m_{B_{u}}^{2})\, V_{ud}V_{cd}^{\ast}\, a_{2}
   \label{a,p-bupi}, \\
   {\cal A}(B_{c}^{+}{\to}B_{u}^{+}{\rho}^{0})
   &=&
   -G_{F}\, F^{B_{c}{\to}B_{u}}_{1}\, f_{\rho}\, m_{\rho}\,
   ({\epsilon}_{\rho}{\cdot}p_{B_{c}})\, V_{ud}V_{cd}^{\ast}\, a_{2}
   \label{amp-burho}, \\
   {\cal A}(B_{c}^{+}{\to}B_{u}^{+}{\omega})
   &=&
   +G_{F}\, F^{B_{c}{\to}B_{u}}_{1}\, f_{\omega}\, m_{\omega}\,
   ({\epsilon}_{\omega}{\cdot}p_{B_{c}})\, V_{ud}V_{cd}^{\ast}\, a_{2}
   \label{amp-buomega}, \\
   {\cal A}(B_{c}^{+}{\to}B_{u}^{+}{\eta}_{q})
   &=&
   -i \frac{G_{F}}{2}\, F^{B_{c}{\to}B_{u}}_{0}\, f_{{\eta}_q}\,
   (m_{B_{c}}^{2}-m_{B_{u}}^{2})\, V_{ud}V_{cd}^{\ast}\, a_{2}
   \label{amp-buetaq}, \\
   {\cal A}(B_{c}^{+}{\to}B_{u}^{+}{\eta}_{s})
  &=&
   -i \frac{G_{F}}{\sqrt{2}}\, F^{B_{c}{\to}B_{u}}_{0}\, f_{{\eta}_s}\,
   (m_{B_{c}}^{2}-m_{B_{u}}^{2})\, V_{us}V_{cs}^{\ast}\, a_{2}
   \label{amp-buetas}, \\
   {\cal A}(B_{c}^{+}{\to}B_{u}^{+}{\eta})
   &=&
   {\cos}{\phi}{\cal A}(B_{c}^{+}{\to}B_{u}^{+}{\eta}_{q})
  -{\sin}{\phi}{\cal A}(B_{c}^{+}{\to}B_{u}^{+}{\eta}_{s})
    \label{amp-bueta}, \\
   {\cal A}(B_{c}^{+}{\to}B_{u}^{+}{\eta}^{\prime})
   &=&
   {\sin}{\phi}{\cal A}(B_{c}^{+}{\to}B_{u}^{+}{\eta}_{q})
  +{\cos}{\phi}{\cal A}(B_{c}^{+}{\to}B_{u}^{+}{\eta}_{s})
    \label{amp-buetap}.
    \end{eqnarray}
  \end{appendix}

  \begin{table}[h]
  \caption{The numerical values of the Wilson coefficients
   and the effective coefficients for $B_{c}$ ${\to}$ $B{\pi}$
   decay, where $m_{c}$ $=$ $1.275{\pm}0.025$ GeV \cite{pdg}.}
  \label{tab02}
  \begin{ruledtabular}
  \begin{tabular}{c|cc|cc|cccc}
 & \multicolumn{2}{c|}{LO} & \multicolumn{2}{c|}{NLO}
 & \multicolumn{4}{c}{QCDF} \\ \cline{2-9}
 ${\mu}$ & $C_{1}$ & $C_{2}$ & $C_{1}$ & $C_{2}$
         & Re($a_{1}$) & Im($a_{1}$) & Re($a_{2}$) & Im($a_{2}$) \\ \hline
 $0.8\,m_{c}$ & $ 1.334$ & $-0.587$ & $ 1.274$ & $-0.503$ & $ 1.270$ & $ 0.096$ & $-0.450$ & $-0.218$ \\
 $     m_{c}$ & $ 1.275$ & $-0.503$ & $ 1.222$ & $-0.424$ & $ 1.216$ & $ 0.068$ & $-0.361$ & $-0.173$ \\
 $1.2\,m_{c}$ & $ 1.239$ & $-0.449$ & $ 1.189$ & $-0.373$ & $ 1.184$ & $ 0.054$ & $-0.306$ & $-0.148$
  \end{tabular}
  \end{ruledtabular}
  \end{table}

  \begin{table}[h]
  \caption{The numerical values of input parameters.}
  \label{input}
  \begin{ruledtabular}
  \begin{tabular}{ll}
  \multicolumn{2}{c}{Wolfenstein parameters} \\ \hline
    ${\lambda}$ $=$ $0.22537{\pm}0.00061$      \cite{pdg}
  & $A$         $=$ $0.814^{+0.023}_{-0.024}$  \cite{pdg} \\
    $\bar{\rho}$ $=$ $0.117{\pm}0.021$         \cite{pdg}
  & $\bar{\eta}$ $=$ $0.353{\pm}0.013$         \cite{pdg} \\ \hline
  \multicolumn{2}{c}{decay constant of mesons} \\ \hline
    $f_{\pi}$      $=$ $130.41{\pm}0.20$ MeV \cite{pdg}
  & $f_{K}  $      $=$ $156.2{\pm}0.7$ MeV \cite{pdg} \\
    $f_{{\eta}_q}$ $=$ $(1.07{\pm}0.02)f_{\pi}$ \cite{prd.58.114006}
  & $f_{{\eta}_s}$ $=$ $(1.34{\pm}0.06)f_{\pi}$ \cite{prd.58.114006} \\
    $f_{\rho}$     $=$ $216{\pm}3$ MeV \cite{ball}
  & $f_{\omega}$   $=$ $187{\pm}5$ MeV \cite{ball} \\
    $f_{K^{\ast}}$ $=$ $220{\pm}5$ MeV \cite{ball} \\ \hline
  \multicolumn{2}{c}{Gegenbauer moments at the scale ${\mu}$ $=$ 1 GeV} \\ \hline
    $a_{1}^{\pi}$ $=$ $a_{1}^{{\eta}_{q}}$ $=$
    $a_{1}^{{\eta}_{s}}$ $=$ $0$ \cite{ball}
  & $a_{2}^{\pi}$ $=$ $a_{2}^{{\eta}_{q}}$ $=$
    $a_{2}^{{\eta}_{s}}$ $=$ $0.25{\pm}0.15$ \cite{ball} \\
    $a_{1}^{\bar{K}}$ $=$ $-a_{1}^{K}$ $=$ $0.06{\pm}0.03$ \cite{ball}
  & $a_{2}^{K}$ $=$ $a_{2}^{\bar{K}}$ $=$ $0.25{\pm}0.15$ \cite{ball} \\
    $a_{1,{\rho}}^{\parallel}$   $=$
    $a_{1,{\omega}}^{\parallel}$ $=$ $0$ \cite{ball}
  & $a_{2,{\rho}}^{\parallel}$   $=$
    $a_{2,{\omega}}^{\parallel}$ $=$ $0.15{\pm}0.07$ \cite{ball} \\
    $a_{1,\bar{K}^{\ast}}^{\parallel}$ $=$
    $-a_{1,K^{\ast}}^{\parallel}$ $=$
    $0.03{\pm}0.02$ \cite{ball}
  & $a_{2,K^{\ast}}^{\parallel}$ $=$
    $a_{2,\bar{K}^{\ast}}^{\parallel}$ $=$ $0.11{\pm}0.09$ \cite{ball} \\
  \end{tabular}
  \end{ruledtabular}
  \end{table}

  \begin{sidewaystable}[h]
  \caption{The $CP$-averaged branching ratios for the $B_{c}$ ${\to}$
   $BP$, $BV$ decays.}
  \label{tab04}
  \begin{ruledtabular}
  {\scriptsize
  \begin{tabular}{l|c|c|c|c|c|c|c|c|c|c|c|c}
    decay mode
  & case
  & Ref. \cite{prd.39.1342}    \footnotemark[1]
  & Ref. \cite{prd.49.3399}    \footnotemark[2]
  & Ref. \cite{prd.62.014019}  \footnotemark[3]
  & Ref. \cite{epjc.32.29}     \footnotemark[4]
  & Ref. \cite{prd.74.074008}  \footnotemark[5]
  & Ref. \cite{prd.80.114003}  \footnotemark[6]
  & Ref. \cite{prd.86.094028}  \footnotemark[7]
  & Ref. \cite{prd.73.054024}  \footnotemark[8]
  & Ref. \cite{pan.67.1559}    \footnotemark[9]
  & Ref. \cite{prd.89.114019}  \footnotemark[10]
  & QCDF
  \\ \hline
    $B_{c}$ ${\to}$ $B_{s}^{0}{\pi}^{+}$
  & 1-a
  & $ 1.0{\times}10^{-1}$
  & $ 6.8{\times}10^{-2}$
  & $ 1.8{\times}10^{-2}$
  & $ 2.9{\times}10^{-2}$
  & $ 4.0{\times}10^{-2}$
  & $ 4.3{\times}10^{-2}$
   ($ 4.3{\times}10^{-2}$)
  & $ 1.3{\times}10^{-1}$
  & $ 4.6{\times}10^{-2}$
  & $ 1.9{\times}10^{-1}$
  & $ 8.8{\times}10^{-2}$
  & $( 1.13^{+0.00+0.11+0.01}_{-0.00-0.06-0.01}){\times}10^{-1}$
 \\
  $B_{c}$ ${\to}$ $B_{s}^{0}K^{+}$
  & 1-b
  & $ 7.6{\times}10^{-3}$
  & $ 4.9{\times}10^{-3}$
  & $ 2.0{\times}10^{-3}$
  & $ 2.4{\times}10^{-3}$
  & $ 3.3{\times}10^{-3}$
  & $ 3.3{\times}10^{-3}$
   ($ 3.3{\times}10^{-3}$)
  & $ 8.5{\times}10^{-3}$
  & $ 3.4{\times}10^{-3}$
  & $ 1.2{\times}10^{-2}$
  & $ 5.2{\times}10^{-3}$
  & $( 7.41^{+0.04+0.70+0.09}_{-0.04-0.39-0.09}){\times}10^{-3}$
  \\
  $B_{c}$ ${\to}$ $B_{s}^{0}{\rho}^{+}$
  & 1-a
  & $ 6.3{\times}10^{-2}$
  & $ 5.2{\times}10^{-2}$
  & $ 4.6{\times}10^{-2}$
  & $ 1.6{\times}10^{-2}$
  & $ 2.7{\times}10^{-2}$
  & $ 3.0{\times}10^{-2}$
   ($ 2.7{\times}10^{-2}$)
  & $ 1.1{\times}10^{-1}$
  & $ 2.7{\times}10^{-2}$
  & $ 8.4{\times}10^{-2}$
  & $ 3.2{\times}10^{-2}$
  & $( 4.44^{+0.00+0.41+0.13}_{-0.00-0.23-0.13}){\times}10^{-2}$
  \\
  $B_{c}$ ${\to}$ $B_{s}^{0}K^{{\ast}+}$
  & 1-b
  & $ 3.2{\times}10^{-4}$
  &
  & $ 1.2{\times}10^{-3}$
  & $ 3.5{\times}10^{-5}$
  & $ 1.5{\times}10^{-4}$
  & $ 8.0{\times}10^{-5}$
   ($ 7.1{\times}10^{-5}$)
  & $ 4.0{\times}10^{-4}$
  & $ 1.3{\times}10^{-4}$
  &
  & $ 9.7{\times}10^{-5}$
  & $( 1.25^{+0.01+0.12+0.06}_{-0.01-0.07-0.06}){\times}10^{-4}$
  \\
  $B_{c}$ ${\to}$ $B_{d}^{0}{\pi}^{0}$
  & 1-b
  & $ 7.4{\times}10^{-3}$
  & $ 3.8{\times}10^{-3}$
  & $ 1.2{\times}10^{-3}$
  & $ 1.2{\times}10^{-3}$
  & $ 1.3{\times}10^{-3}$
  & $ 1.8{\times}10^{-3}$
   ($ 1.5{\times}10^{-3}$)
  & $ 8.4{\times}10^{-3}$
  & $ 2.4{\times}10^{-3}$
  & $ 1.2{\times}10^{-2}$
  & $ 6.9{\times}10^{-3}$
  & $( 7.83^{+0.04+0.73+0.04}_{-0.04-0.41-0.04}){\times}10^{-3}$
  \\
  $B_{c}$ ${\to}$ $B_{d}^{0}K^{+}$
  & 1-c
  &
  & $ 3.0{\times}10^{-4}$
  & $ 1.2{\times}10^{-4}$
  & $ 1.0{\times}10^{-4}$
  & $ 1.1{\times}10^{-4}$
  & $ 1.5{\times}10^{-4}$
   ($ 1.2{\times}10^{-4}$)
  & $ 5.9{\times}10^{-4}$
  & $ 1.8{\times}10^{-4}$
  & $ 8.1{\times}10^{-4}$
  & $ 4.4{\times}10^{-4}$
  & $( 5.29^{+0.06+0.50+0.07}_{-0.06-0.28-0.07}){\times}10^{-4}$
  \\
  $B_{c}$ ${\to}$ $B_{d}^{0}{\rho}^{+}$
  & 1-b
  & $ 8.3{\times}10^{-3}$
  & $ 6.9{\times}10^{-3}$
  & $ 3.3{\times}10^{-3}$
  & $ 1.5{\times}10^{-3}$
  & $ 1.6{\times}10^{-3}$
  & $ 2.2{\times}10^{-3}$
   ($ 1.7{\times}10^{-3}$)
  & $ 1.4{\times}10^{-2}$
  & $ 2.4{\times}10^{-3}$
  & $ 1.1{\times}10^{-2}$
  & $ 4.3{\times}10^{-3}$
  & $( 5.32^{+0.03+0.49+0.15}_{-0.03-0.28-0.15}){\times}10^{-3}$
  \\
  $B_{c}$ ${\to}$ $B_{d}^{0}K^{{\ast}+}$
  & 1-c
  &
  & $ 2.1{\times}10^{-4}$
  & $ 1.5{\times}10^{-4}$
  & $ 4.6{\times}10^{-5}$
  & $ 4.4{\times}10^{-5}$
  & $ 4.9{\times}10^{-5}$
   ($ 3.7{\times}10^{-5}$)
  & $ 3.5{\times}10^{-4}$
  & $ 5.7{\times}10^{-5}$
  & $ 1.7{\times}10^{-4}$
  & $ 8.3{\times}10^{-5}$
  & $( 1.06^{+0.01+0.10+0.05}_{-0.01-0.06-0.05}){\times}10^{-4}$
  \\
  $B_{c}$ ${\to}$ $B_{u}^{+}\overline{K}^{0}$
  & 2-a
  & $ 2.1{\times}10^{-2}$
  & $ 1.2{\times}10^{-2}$
  & $ 4.9{\times}10^{-3}$
  & $ 4.2{\times}10^{-3}$
  & $ 4.4{\times}10^{-3}$
  & $ 6.0{\times}10^{-3}$
   ($ 4.9{\times}10^{-3}$)
  & $ 2.3{\times}10^{-2}$
  & $ 6.8{\times}10^{-3}$
  & $ 3.6{\times}10^{-2}$
  & $ 2.2{\times}10^{-3}$
  & $( 1.97^{+0.00+1.11+0.05}_{-0.00-0.54-0.05}){\times}10^{-2}$
  \\
  $B_{c}$ ${\to}$ $B_{u}^{+}K^{0}$
  & 2-c
  &
  &
  &
  &
  &
  & $ 1.6{\times}10^{-5}$
   ($ 1.3{\times}10^{-5}$)
  &
  &
  &
  & $ 6.3{\times}10^{-6}$
  & $( 5.71^{+0.06+3.20+0.15}_{-0.06-1.58-0.14}){\times}10^{-5}$
  \\
  $B_{c}$ ${\to}$ $B_{u}^{+}\overline{K}^{{\ast}0}$
  & 2-a
  & $ 7.8{\times}10^{-3}$
  & $ 8.5{\times}10^{-3}$
  & $ 5.8{\times}10^{-3}$
  & $ 1.6{\times}10^{-3}$
  & $ 1.6{\times}10^{-3}$
  & $ 1.9{\times}10^{-3}$
   ($ 1.4{\times}10^{-3}$)
  & $ 1.3{\times}10^{-2}$
  & $ 2.1{\times}10^{-3}$
  & $ 8.0{\times}10^{-3}$
  & $ 2.0{\times}10^{-4}$
  & $( 3.72^{+0.00+2.09+0.21}_{-0.00-1.03-0.20}){\times}10^{-3}$
  \\
  $B_{c}$ ${\to}$ $B_{u}^{+}K^{{\ast}0}$
  & 2-c
  &
  &
  &
  &
  &
  & $ 5.0{\times}10^{-6}$
   ($ 3.7{\times}10^{-6}$)
  &
  &
  &
  & $ 5.6{\times}10^{-7}$
  & $( 1.07^{+0.01+0.60+0.06}_{-0.01-0.30-0.06}){\times}10^{-5}$
  \\
  $B_{c}$ ${\to}$ $B_{u}^{+}{\pi}^{0}$
  & 2-b
  & $ 4.0{\times}10^{-4}$
  & $ 2.1{\times}10^{-4}$
  & $ 6.4{\times}10^{-5}$
  & $ 6.2{\times}10^{-5}$
  & $ 6.7{\times}10^{-5}$
  & $ 9.7{\times}10^{-5}$
   ($ 8.1{\times}10^{-5}$)
  & $ 4.5{\times}10^{-4}$
  & $ 1.3{\times}10^{-4}$
  & $ 6.6{\times}10^{-4}$
  & $ 5.2{\times}10^{-5}$
  & $( 4.23^{+0.02+2.37+0.07}_{-0.02-1.17-0.07}){\times}10^{-4}$
 \\
  $B_{c}$ ${\to}$ $B_{u}^{+}{\rho}^{0}$
  & 2-b
  & $ 4.4{\times}10^{-4}$
  & $ 3.7{\times}10^{-4}$
  & $ 1.7{\times}10^{-4}$
  & $ 8.7{\times}10^{-5}$
  & $ 8.9{\times}10^{-5}$
  & $ 1.2{\times}10^{-4}$
   ($ 9.4{\times}10^{-5}$)
  & $ 7.4{\times}10^{-4}$
  & $ 1.3{\times}10^{-4}$
  & $ 5.5{\times}10^{-4}$
  & $ 1.8{\times}10^{-5}$
  & $( 2.86^{+0.01+1.60+0.10}_{-0.01-0.79-0.10}){\times}10^{-4}$
  \\
  $B_{c}$ ${\to}$ $B_{u}^{+}{\omega}$
  & 2-b
  & $ 4.1{\times}10^{-4}$
  &
  &
  &
  &
  & $ 9.0{\times}10^{-5}$
   ($ 7.0{\times}10^{-5}$)
  &
  &
  &
  & $ 1.3{\times}10^{-5}$
  & $( 2.05^{+0.01+1.15+0.12}_{-0.01-0.57-0.12}){\times}10^{-4}$
  \\
  $B_{c}$ ${\to}$ $B_{u}^{+}{\eta}$
  &
  &
  &
  &
  &
  &
  & $ 5.0{\times}10^{-4}$
   ($ 4.1{\times}10^{-4}$)
  &
  &
  &
  & $ 1.4{\times}10^{-4}$
  & $( 1.46^{+0.01+0.82+0.13}_{-0.01-0.40-0.12}){\times}10^{-3}$
  \\
  $B_{c}$ ${\to}$ $B_{u}^{+}{\eta}^{\prime}$
  &
  &
  &
  &
  &
  &
  & $ 6.7{\times}10^{-6}$
   ($ 5.6{\times}10^{-6}$)
  &
  &
  &
  & $ 4.2{\times}10^{-6}$
  & $( 7.28^{+0.04+4.09+1.66}_{-0.04-2.01-1.47}){\times}10^{-5}$
  \end{tabular}
  }
  \footnotetext[1]{It is estimated with the form factors
     $F_{0}^{B_{c}{\to}B_{s}}$ $=$ $0.925$,
     $F_{0}^{B_{c}{\to}B}$ $=$ $0.91$ and
     parameter ${\omega}$ $=$ 1 GeV \cite{prd.39.1342}
     based on the BSW model.}
  \footnotetext[2]{It is estimated with the instantaneous
     nonrelativistic approximation and the potential model
     based on the BS equation.}
  \footnotetext[3]{It is estimated in a relativistic model
     with a one-gluon interaction plus a scalar confinement
     potential based on the BS equation.}
  \footnotetext[4]{It is estimated with the form factors
     $F_{0}^{B_{c}{\to}B_{s}}$ $=$ $0.5$,
     $F_{0}^{B_{c}{\to}B}$ $=$ $0.39$ using a
     quasipotential in the relativistic quark model \cite{epjc.32.29}.}
  \footnotetext[5]{It is estimated with the form factors
     $F_{0}^{B_{c}{\to}B_{s}}$ $=$ $0.58$,
     $F_{0}^{B_{c}{\to}B}$ $=$ $0.39$ in the
     nonrelativistic constituent quark model \cite{prd.74.074008}.}
  \footnotetext[6]{It is estimated with the form factors
     $F_{0}^{B_{c}{\to}B_{s}}$ $=$ $0.573\,(0.571)$,
     $F_{0}^{B_{c}{\to}B}$ $=$ $0.467\,(0.426)$ in the
     light-front quark model based on the Coulomb plus linear
     (harmonic oscillator) potential, together with the hyperfine
     interaction \cite{prd.80.114003}.}
  \footnotetext[7]{It is estimated with the form factors
     $F_{0}^{B_{c}{\to}B_{s}}$ $=$ $1.03$,
     $F_{0}^{B_{c}{\to}B}$ $=$ $1.01$ in the
     relativistic independent quark model \cite{prd.86.094028}.}
  \footnotetext[8]{It is estimated within a relativistic constituent
     quark model \cite{prd.73.054024}.}
  \footnotetext[9]{It is estimated with the form factors
     $F_{0}^{B_{c}{\to}B_{s}}$ $=$ $1.3$,
     $F_{0}^{B_{c}{\to}B}$ $=$ $1.27$ in the
     QCD sum rules \cite{pan.67.1559}.}
  \footnotetext[10]{It is estimated with the perturbative QCD approach
     based on the $k_{T}$ factorization scheme \cite{prd.89.114019}.}
  \end{ruledtabular}
  \end{sidewaystable}

\begin{thebibliography}{99}
  \bibitem{pdg}
      K. Olive {\em et al}. (Particle Data Group),
      Chin. Phys. C 38, 090001 (2014).
  \bibitem{qwg}
     N. Brambilla {\em et al}. (Quarkonium Working Group),
     arXiv:hep-ph/0412158, and references therein.
  \bibitem{1st}
     F. Abe {\em et al}. (CDF Collaboration),
     Phys. Rev. D 58, 112004 (1998);
     Phys. Rev. Lett. 81, 2432 (1998).
  \bibitem{mass}
     R. Aaij {\em et al}. (LHCb Collaboration),
     Phys. Rev. D 87, 112012 (2013).
  \bibitem{time}
     R. Aaij {\em et al}. (LHCb Collaboration),
     Phys. Lett. B 742, 29 (2015).
  \bibitem{events}
     I. Gouz {\em et al.},
     Phys. Atom. Nucl. 67, 1559 (2004);
     A. Likhoded, A. Luchinsky,
     Phys. Rev. D 82, 014012 (2010).
  \bibitem{zpc51}
     M. Lusignoli, M. Masetti,
     Z. Phys. C 51, 549 (1991).
  \bibitem{prd49}
     C. Chang, Y. Chen,
     Phys. Rev. D 49, 3399 (1994).
  \bibitem{usp38}
     S. Gershtein {\em et al.},
     Phys. Usp. 38, 1 (1995).
  \bibitem{bspi}
     R. Aaij {\em et al}. (LHCb collaboration),
     Phys. Rev. Lett. 111, 181801 (2013).
  \bibitem{prd53}
     M. Beneke and G. Buchalla,
     Phys. Rev. D 53, 4991 (1996);
     C. Chang, S. Chen, T. Feng and X. Li,
     Phys. Rev. D 64, 014003 (2001);
     Commun. Theor. Phys. 35, 57 (2001).
  \bibitem{prd.39.1342}
     D. Du and Z. Wang,
     Phys. Rev. D 39, 1342 (1989).
  \bibitem{zpc.51.549}
     M. Lusignoil and M. Masetti,
     Z. Phys. C 51, 549 (1991).
  \bibitem{pan.62.1793}
     A. Anisimov, P. Kulikov, I. Narodetskii and K. Ter-Martirosyan,
     Phys. Atom. Nucl. 62, 1739 (1999).
  \bibitem{jpg.35.085002}
     R. Dhir, N. Sharma and R. Verma,
     J. Phys. G 35, 085002 (2008);
     R. Dhir and R. Verma,
     Phys. Rev. D 79, 034004 (2009).
  \bibitem{bsw}
     M. Wirbel, B. Stech and M. Bauer,
     Z. Phys. C 29, 637 (1985);
     M. Bauer, B. Stech and M. Wirbel,
     Z. Phys. C 34, 103 (1987).
  \bibitem{igsw}
     N. Isgur, D. Scora, B. Grinstein and N. Wise,
     Phys. Rev. D 39, 799 (1989).
  \bibitem{prd.49.3399}
     C. Chang and Y. Chen,
     Phys. Rev. D 49, 3399 (1994).
  \bibitem{prd.56.4133}
     J. Liu and K. Chao,
     Phys. Rev. D 56, 4133 (1997).
  \bibitem{prd.62.014019}
     A. El-Hady, J. Mu\~{n}oz and J. Vary,
     Phys. Rev. D 62, 014019 (2000).
  \bibitem{prd.61.034012}
     P. Colangelo and F. Fazio,
     Phys. Rev. D61, 034012 (2000).
  \bibitem{epjc.32.29}
     D. Ebert, R. Faustov and V. Galkin,
     Eur. Phys. J. C 32, 29 (2003).
  \bibitem{prd.74.074008}
     E. Hern\'{a}ndez, J. Nieves and J. Verde-Velasco,
     Phys. Rev. D 74, 074008 (2006);
     Eur. Phys. J. A 31, 714 (2007).
  \bibitem{prd.80.114003}
     H. Choi and C. Ji,
     Phys. Rev. D 80, 114003 (2009).
  \bibitem{prd.86.094028}
     Sk. Naimuddin {\em et al.},
     Phys. Rev. D 86, 094028 (2012).
  \bibitem{prd.73.054024}
     M. Ivanov, J. K\"{o}rner and P. Santorelli,
     Phys. Rev. D 73, 054024 (2006).
  \bibitem{pan.67.1559}
     V. Kiselev, A. Kovalsky and A. Likhoded,
     Nucl. Phys. B 585, 353 (2000);
     Phys. Atom. Nucl. 64, 1860 (2001);
     I. Gouz {\em et al.},
     Phys. Atom. Nucl. 67, 1559 (2004).
  \bibitem{prd.65.114007}
     R. Verma and A. Sharma,
     Phys. Rev. D 65, 114007 (2002).
  \bibitem{prd.89.114019}
     J. Sun, Y. Yang, Q. Chang and G. Lu,
     Phys. Rev. D 89, 114019 (2014);
     J. Sun, Y. Yang and G. Lu,
     Sci. China Phys. Mech. Astron. 57, 1891 (2014).
  \bibitem{pqcd}
     C. Chang and  H. Li,
     Phys. Rev. D 55, 5577 (1997);
     T. Yeh and H. Li,
     Phys. Rev. D 56, 1615 (1997);
     Y. Keum, H. Li and A. Sanda,
     Phys. Lett. B 504, 6 (2001);
     Y. Keum and H. Li,
     Phys. Rev. D 63, 074006 (2001);
     C. L\"{u}, K. Ukai and M. Yang,
     Phys. Rev. D 63, 074009 (2001);
     C. L\"{u} and M. Yang,
     Eur. Phys. J. C 23, 275 (2002).
  \bibitem{scet}
      C. Bauer, S. Fleming and M. Luke,
      Phys. Rev. D 63, 014006 (2001);
      C. Bauer et al.,
      Phys. Rev. D 63, 114020 (2001);
      C. Bauer and I. Stewart,
      Phys. Lett. B 516, 134 (2001);
      C. Bauer, D. Pirjol and I. Stewart,
      Phys. Rev. D 65, 054022 (2002);
      C. Bauer, et al.,
      Phys. Rev. D 66, 014017 (2002);
      M. Beneke et al.,
      Nucl. Phys. B 643, 431 (2002);
      M. Beneke and T. Feldmann,
      Phys. Lett. B 553, 267 (2003);
      Nucl. Phys. B 685, 249 (2004).
  \bibitem{qcdf}
      M. Beneke et al.,
      Phys. Rev. Lett. 83, 1914 (1999);
      Nucl. Phys. B 591, 313 (2000);
      Nucl. Phys. B 606, 245 (2001);
      D. Du, D. Yang, G. Zhu,
      Phys. Lett. B 488, 46 (2000);
      Phys. Lett. B 509, 263 (2001);
      Phys. Rev. D 64, 014036 (2001).
  \bibitem{9512380}
      For a review, see
      G. Buchalla, A. Buras  and M. Lautenbacher,
      Rev. Mod. Phys. 68, 1125, (1996)
      or A. Buras, arXiv:hep-ph/9806471.
  \bibitem{du}
      D. Du, H, Gong, J. Sun, D. Yang, and G. Zhu,
      Phys. Rev. D 65, 074001 (2002); 
      Phys. Rev. D 65, 094025 (2002);
      Erratum, ibid. D 66, 079904 (2002); 
      J. Sun, G. Zhu and D. Du,
      Phys. Rev. D 68, 054003 (2003). 
  \bibitem{beneke}
      M. Beneke and M. Neubert,
      Nucl. Phys. B 675, 333 (2003);
      M. Beneke, J. Rohrer and D. Yang,
      Nucl. Phys. B 774, 64 (2007).
  \bibitem{prd.73.114027}
      X. Li and Y. Yang,
      Phys. Rev. D 73, 114027 (2006).
  \bibitem{cheng}
      H. Cheng and K. Yang,
      Phys. Rev. D 78, 094001 (2008);
      Erratum, ibid. D 79, 039903 (2009);
      H. Cheng and C. Chua,
      Phys. Rev. D 80, 074031 (2009); 
      Phys. Rev. D 80, 114008 (2009); 
      Phys. Rev. D 80, 114026 (2009); 
      Phys. Rev. D 83, 034001 (2011). 
  \bibitem{ball}
      P. Ball, JHEP 01, 010 (1999);
      P. Ball, V. Braun and A. Lenz, JHEP 05, 004 (2006);
      P. Ball and G. Jones, JHEP 03, 069 (2007).
  \bibitem{prd.58.114006}
     Th. Feldmann, P. Kroll and B. Stech,
     Phys. Rev. D 58, 114006 (1998).
  \bibitem{scale}
     G. Bell, Nucl. Phys. B 822, 172 (2009);
     M. Beneke, T. Huber, X. Li, Nucl. Phys. B 832, 109 (2010).
  \end{thebibliography}
  \end{document}